\documentclass[final,5p,times,twocolumn]{elsarticle}

 \usepackage{graphicx}
\usepackage{color}
\usepackage{amssymb}

\def\beq{\begin{equation}}
\def\eeq{\end{equation}}
\def\bea{\begin{eqnarray}}
\def\eea{\end{eqnarray}}
\def\vd{\langle v_d q \rangle}

\newcommand*{\eqref}[1]{Eq.~(\ref{eq:#1})}
\newcommand*{\eqlab}[1]{\label{eq:#1}}
\newcommand*{\figref}[1]{Fig.~\ref{fig:#1}}
\newcommand*{\figlab}[1]{\label{fig:#1}}

\newcommand*{\secref}[1]{Section~\ref{sec:#1}}
\newcommand*{\seclab}[1]{\label{sec:#1}}

\def\VYP#1#2#3{{\bf #1}, #3 (#2)}  

\def\PL#1#2#3{Phys.~Lett.~\VYP{#1}{#2}{#3}}

\newcommand{\etal}{\mbox{\textit et al.}}                       %
\newcommand{\Omit}[1]{}

\journal{Nuclear Instruments and Methods in Physics Research, A}

\begin{document}

\begin{frontmatter}



\title{Macroscopic Model of Geomagnetic-Radiation from Air Showers II.}


\author[KVI]{Olaf Scholten}
\ead{scholten@kvi.nl}
\author[KVI]{Krijn D. de Vries}
\author[SUBA]{Klaus Werner}
\address[KVI]{Kernfysisch Versneller Instituut, University of Groningen, 9747 AA, Groningen, The Netherlands}
\address[SUBA]{SUBATECH, University of Nantes -- IN2P3/CNRS-- EMN,  Nantes, France}

\begin{abstract}
The generic properties of the emission of coherent radiation from a moving charge distribution are discussed. The general structure of the charge and current distributions in an extensive air shower are derived. These are subsequently used to develop a very intuitive picture for the properties of the emitted radio pulse. Using this picture can be seen that the structure of the pulse is a direct reflection of the shower profile. At higher frequencies the emission is suppressed because the wavelength is shorter than the important length scale in the shower. It is shown that radio emission can be used to distinguish proton and iron induced air showers.
\end{abstract}

\end{frontmatter}


\section{Introduction}

In recent years the field of radio detection of cosmic ray air showers has reached a mature stage which is shown by the many  contributions to this meeting on the topic of radio emission. Consensus is approaching on the theoretical description~\cite{Tim} of the emission process and a detailed quantitative understanding of the experimental results~\cite{Marianne, Harm} is close although some challenges remain~\cite{Krijn}.
Installation is in progress of extensive arrays of radio detectors at the Pierre Auger Observatory~\cite{RAuger}, at LOFAR~\cite{Andreas}, and at the South pole~\cite{Boeser}.
In this work the importance of coherent radio emission from air showers will be stressed where the most important emission mechanisms were already investigated in the earliest works on this subject~\cite{Jel65,Por65,Kah66,All71}, namely Cherenkov and geo-magnetic radiation. A complete historical review is given in Ref.~\cite{History}.

Coherent emission occurs when the emitting charges are confined to distances $\Lambda$ which are smaller than the wavelength $\lambda$. Since the source is `viewed' with a resolution of the wavelength, a fine sub-structure in the source will not affect the emission process.  At a much shorter wavelength, $\lambda \ll \Lambda$, the different parts of the fine structure in the charge distribution will contribute to the emission process with varying phases which as often result in constructive as in destructive interference with as result that the net emission probability is strongly suppressed.
In an EAS initiated by a cosmic ray of $10^{18}$\,eV the number of charged particles at the shower maximum is of the order of $N=10^6$ and coherent radiation, where the intensity is proportional to $N^2$, is far more intense that incoherent radiation where the intensity is proportional to $N$. Only at high frequencies, where the coherent process is suppressed because the wave length is much smaller that the relevant size of the emitting charge distribution, the incoherent process can contribute.
For coherent emission thus only the macroscopic structures in the EAS contribute.

At high frequency the coherence condition $\Lambda < \lambda$ will no longer be satisfied implying a cut-off of the coherent response.
Some of the length scales that are important for EAS emission~\cite{Arena08} are i) the pancake thickness; ii) the length of the EAS projected along the line of sight; iii) the lateral distribution of the charges in the EAS.
The high-frequency cut-off is reflected in the time between the start and the zero crossing of the pulse. The important aspect of the macroscopic model is that it clearly shows which are the important lengths that determine the emission process.

Model independent conclusions can also be drawn at large wave-length since the intensity of the emitted radiation of a
system of charges decreases linearly with increasing wave-length when considerably larger than the the size of the emitting body.
An EAS, independent how enormous the event may be, always exists for a limited
time and occurs in a limited part of the atmosphere and currents are confined to a limited region of space-time. As a consequence  the intensity of the emitted radio waves should vanish linearly in the limit of infinite wave length or zero frequency.
The time integral of the emitted pulse should vanish implying equally large positive and negative amplitudes. The simplest structure of a pulse is thus bi-polar and unipolar pulses are un physical~\cite{Arena08}.

For a correct description it is important to have a consistent description of the charges and their motion. Consistent in this respect means that all charges are accounted for and thus charge conservation holds. Charges may move and be suddenly accelerated (through a collision) but no net charge can be created. For example, if an electron is accelerated in a Compton scattering process, a positively charged ion should remain behind. Once the consistent description is obtained of the distribution of the charges and their velocities, resulting in a four-current density $j^\mu(\vec{r},t)$, the emitted radiation is straightforward to calculate by applying the Maxwell equations with the current density as input. As mentioned, for coherent emission only densities averaged over an appropriately chosen length scale matter where one should be careful to account for {\em all } charges and currents in the system to obey charge conservation. Due to the necessary averaging over the path of many individual electrons a very simple picture emerges for the emission process which lies at the basis of the Macroscopic GeoMagnetic Radiation (MGMR) model~\cite{Sch08,Wer08}.

In \secref{MGMR} the essential current distributions for a generic EAS are derived which form the basis for the (semi) analytic predictions for the structure of the radio pulse in the MGMR model~\cite{Sch08,Wer08}. As an application we show, using a hybrid approach where the parameters of the MGMR model are extracted from a Monte-Carlo simulation of the air-shower, that the radio emission can be used to distinguish proton and iron induced showers~\cite{Krijn}.

\section{The Macroscopic Model for Geo-Magnetic Radiation\seclab{MGMR}}

The pancake at the front of an EAS consists out of a plasma with large amounts of electrons, positrons and other particles moving towards the surface of the Earth with a velocity almost equal to the light velocity.
In the center of the shower constantly electron-positron pairs are created, mainly by the energetic photons in the shower core, to form the pancake, depicted by the broad band in \figref{MGMR}. The shower front, as it is driven by photons, moves towards the Earth with the light velocity while the produced particles trail some distance behind the front and are responsible for the finite thickness of the pancake. A net electric current is induced in the pancake because of the Earth's magnetic field induces a Lorentz force pulling the electrons and positrons in opposite direction. Because of the constant interactions with the air molecules the leptons (electrons and positrons) reach a constant drift velocity $v_d$ as corroborated by Monte-Carlo shower calculations~\cite{Krijn}. In \figref{MGMR} the drifting electrons and positrons are denoted by the chequered arrows. The motion of the electrons and the positrons thus contribute coherently to a net electric current density in the direction indicated by the red arrow. Due to the constant interaction with the ambient air molecules the leptons will loose energy and trail further behind the shower front and become non-relativistic. In the figure these leptons are denoted as `stopped'. On average the electrons and positrons are separated by a distance $D_S$. The fact that this happens can also be seen as a consequence of charge conservation, the electric current must induce a displacement of net charge from one side of the shower core to the other. One consequence of the induced current density may not be directly obvious. As the electrons and positrons move in opposite directions there is already a net displacement of charges that move with relativistic velocities. Their displacement is about half of the stopped charges, $D_M\approx D_S/2$. This completes the qualitative picture of the charges and currents that are induced by the geomagnetic field. For the following discussion it is instructive to distinguish the different components which must be present in any realistic shower simulation.

\begin{figure}[ht]
 \centerline{ \includegraphics[width=0.47\textwidth]{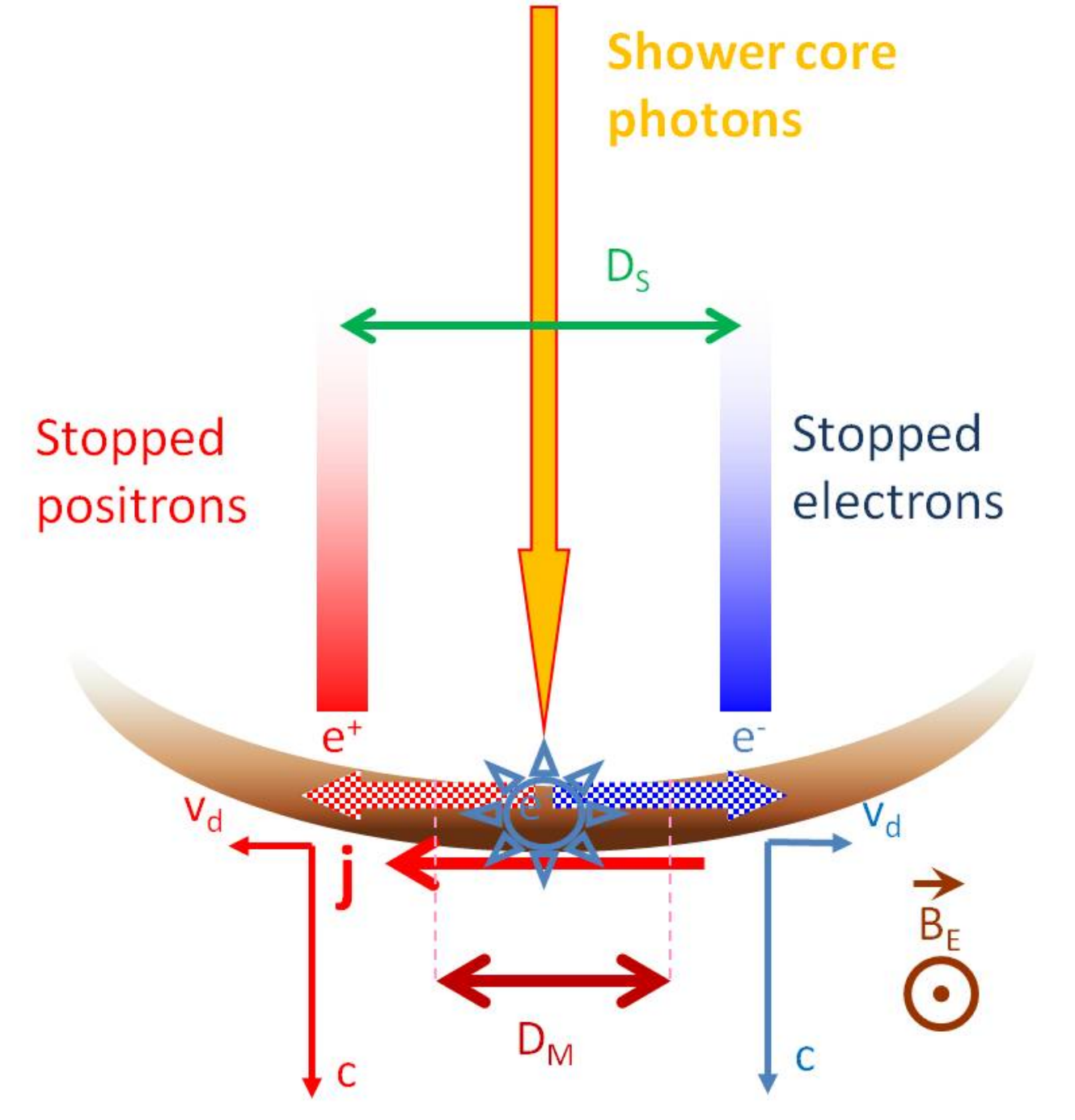} }
\caption[fig1-MGMR]{[color online]
Schematic description of the current densities in an Extensive Air Shower~\cite{Sch08}.}
  \figlab{MGMR}
\end{figure}

In addition to the current and charge densities of geomagnetic origin there is also an induced charge that is independent of the magnetic field. Through the process of Compton scattering on electrons in air molecules and electron knock-out reactions by the relativistic electrons a net excess of relativistic electrons is created in the pancake. Simulations indicate that this excess can be large, about 30\% of the total lepton density~\cite{Krijn} and this is indicated in \figref{MGMR} by the star in the shower core. The net negative charge excess implies that there must also be a positive charge in the system, which is formed by positively charged air molecules. Even though these positive charges are at rest, they contribute to the radiation field since their number increases as the shower develops, but are not marked in \figref{MGMR} for simplicity.

To obtain better insight in the emission mechanism we will use the picture of the currents described in \figref{MGMR}, realizing that all charges are concentrated very near the shower axis. As is discussed in detail in Ref.~\cite{Sch08,Wer08} this results in a picture where the geomagnetic electric current becomes point-like at the shower axis and the charge separation of the relativistic leptons is taken into account as an electric dipole moving at a relativistic speed. The stopped particles are similarly accounted for by an electric dipole that is at rest in the Earth frame but constantly increasing in magnitude as the shower front proceeds to the surface of the Earth. As shown~\cite{Sch08,Wer08}, the induced electric current and the net charge excess give the dominant contribution to the emitted radiation.

To obtain an analytic expression for the pulse due to the geomagnetic current the thickness of the pancake is assumed to be small~\cite{Sch08} and we adopt a simple geometry with a vertical shower and a horizontal magnetic field.  The number
of electrons and positrons in the shower at an height $z=-c t_r$ is parameterized as $N(z)=N_e f_t(t_r)$ in terms of the normalized shower profile, $f_t(t_r)$, where $N_e$ is the number of electrons in the shower at the maximum. The induced geomagnetic current in the $\hat{x}$-direction is
\beq
 j(x,y,z,t)= \vd \, e\, N_e f_t(t_r) \;.
\eqlab{CurrDens}
\eeq
In the simple picture the drift velocity is assumed to be independent on the height in the atmosphere.
In the limit where the shower moves with the light velocity and the index of refraction of air equals unity, the retarded time $t_r$ can be expressed in terms of the observer time $t$ as
\beq
ct_r \approx -{d^2 \over 2 c t}  \;.
 \eqlab{t-ret-app}
\eeq
The observer is at a distance $d$ from the core and $t$ is time after the shower hits the surface of the Earth.
\eqref{t-ret-app} shows that the early part of the received pulse is emitted at large (and negative) retarded times and thus large heights while the late part of the pulse is emitted when the EAS was already close to the round.

The only non-vanishing component of the vector potential is in the direction of the electric current,
\beq
A^x(t,d)= J {f_t(t_r) \over {\cal D}} \;,\eqlab{Ax}
\eeq
where $J= \vd N_e e/ 4 \pi \varepsilon_0 c$ is a constant depending on the energy of the cosmic ray and ${\cal D}$ is the retarded distance. The electric field is proportional to the time derivative of $\vec{A}$, giving
\beq
 E_x(t,d) \approx J {c^2 t_r^2 4 \over d^4} {d\over dt_r}[ t_r f_t(t_r)] \;,
\eqlab{E-appx}
\eeq
using \eqref{t-ret-app}.
From this simple expression some important observations can be made which also hold for realistic calculations to a surprisingly accuracy. The first is that radiation is emitted due to the variation of the shower profile with height.
The zero crossing of the electric field at the observer position corresponds to the maximum in $t_r f_t(t_r)$ and thus the emission from the shower at an height exceeding that of the shower maximum. The dominant part of the
pulse is thus emitted at heights well above the shower maximum which implies that
the radio signal bears information on the early stages of the EAS development.
A second observation is that, because the time dependence of the pulse is expressed in terms of the shower time $t_r$ \eqref{t-ret-app}, it can be seen that at twice the distance from the point of impact the pulse is four times as wide while the amplitude is decreased by a factor $2^4$. It can easily be seen that this simple picture is not valid at small distances to the shower core where the pancake thickness becomes the determining length scale.

\begin{figure}[hhc!]
\centerline{
\includegraphics[width=.48\textwidth,keepaspectratio]{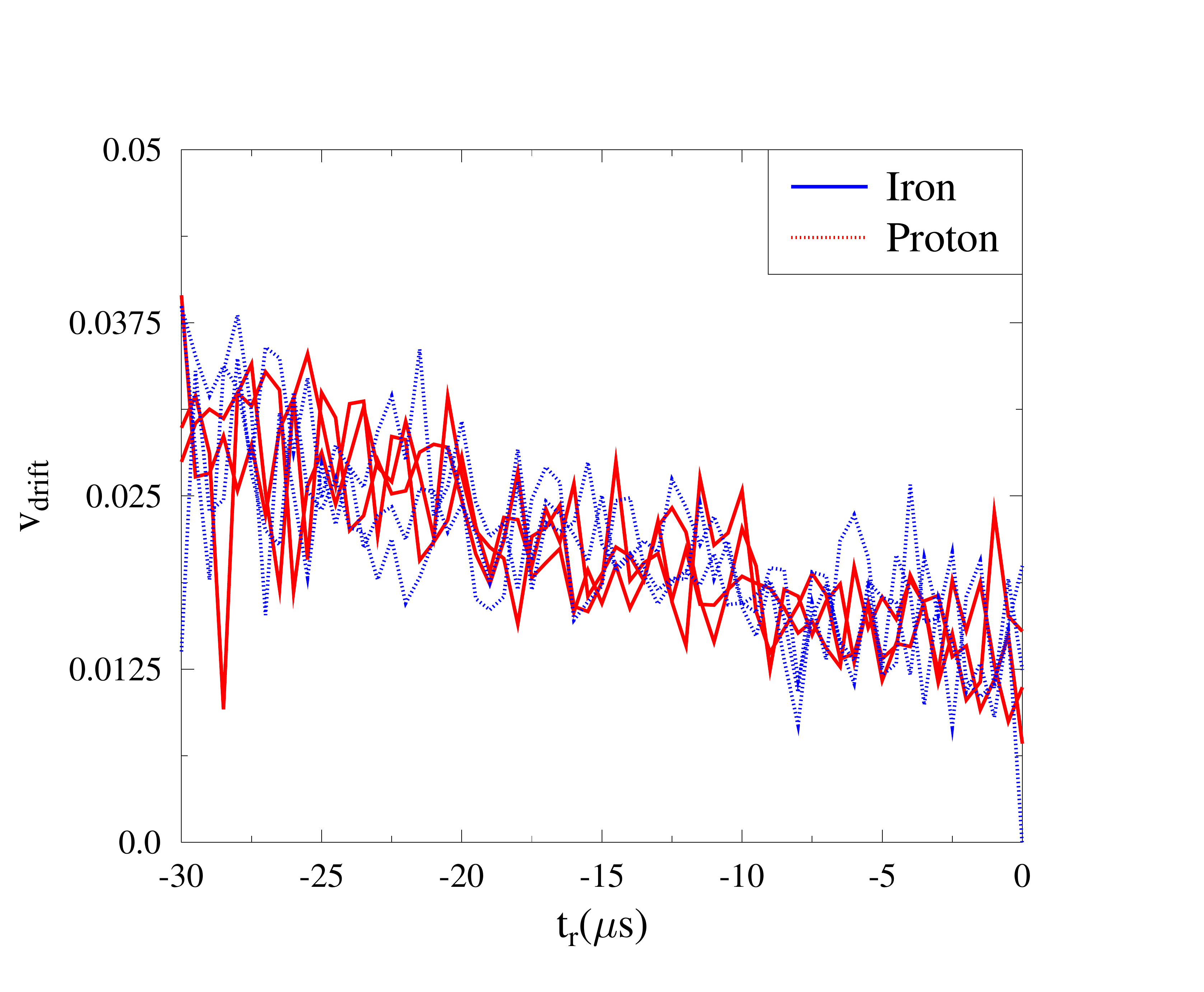}
}
\caption{[color online] Drift velocities as function of shower time $t_r$ for a few proton (blue curves) and iron (red curves) initiated showers. }
\figlab{DriftVel}
\end{figure}

A prime example showing the strength of the macroscopic approach is the introduction of the drift velocity. The motion of any given electron (or positron) in the shower is very complicated. It is suddenly accelerated (or created) and moves at relativistic velocities in the Earth magnetic field where it experiences a constant acceleration due to the action of the Lorentz force. Through soft processes, mainly Thompson scattering, it is constantly loosing energy and decelerating. It may be strongly deflected in hard scattering processes. With all acceleration processes radiation will be emitted. For coherent emission the motion of all electrons in a certain phase-space element has to be averaged. At any time some of the electrons in a volume will be accelerated by the Lorentz force while others are decelerated by the action of the air. The net result is that on average the electrons move with a constant drift velocity in one direction and the positrons in the other. This picture is supported by Monte-Carlo calculations showing that the drift velocity depends on air density as shown in \figref{DriftVel}. As a result of reaching an equilibrium with an (almost) constant drift velocity neither synchrotron emission (acceleration) nor bremsstrahlung (deceleration) is of particular importance and only the time variation of the induced macroscopic electric current causes the EAS to emit radiation.

In the complete MGMR calculations a 4-current, $j^{\mu}(\vec{r},t)$, is constructed where the zeroth component corresponds to the charge and the other components to the electric current. The electric field at the observer position $d$ and time $t$ can be expressed as a derivative of the Li\'{e}nard-Wiechert potential which in turn can be calculated directly from the 4-current density following any standard textbook on Electromagnetism~\cite{Jac-CE}
\begin{equation}
 A^{\mu}(t,d)= \frac{1}{4 \pi \varepsilon_0} \int
 \left. {j^\mu\over R(1-\vec{\beta}\cdot \hat{n})}\right|_{\mbox{ret}} \,dh\;,
 \eqlab{LW}
\end{equation}
ignoring the lateral extension of the shower profile and where the integration runs over the pancake thickness $h$.  We use the common notation where $\hat{n}$ is a unit vector pointing from the source to the
observer and $R$ is the distance, both evaluated at the retarded time.

\section{Distance dependence of the radio pulse \& composition}

Up to this point we have focussed on obtaining a qualitative description of the current densities in an EAS. To make quantitative comparisons with data a hybrid approach has been developed where the important model parameters are obtained from Monte-Carlo simulations~\cite{Krijn}.

These simulations are done using the cascade mode of the CONEX shower simulation program (where further details can be found in Ref.~\cite{Krijn}) modified to include the deviation of charged particles in the Earth's magnetic field~\cite{Wer08}.
An analysis tool has been written to give the full three-dimensional and timing information of the currents and particle distributions in histograms.
More details of the complete shower-simulation package including analysis tools, called CONEX-MC-GEO, will be discussed in a future publication.

\begin{figure}[hhc!]
\centerline{
\includegraphics[width=.48\textwidth,keepaspectratio]{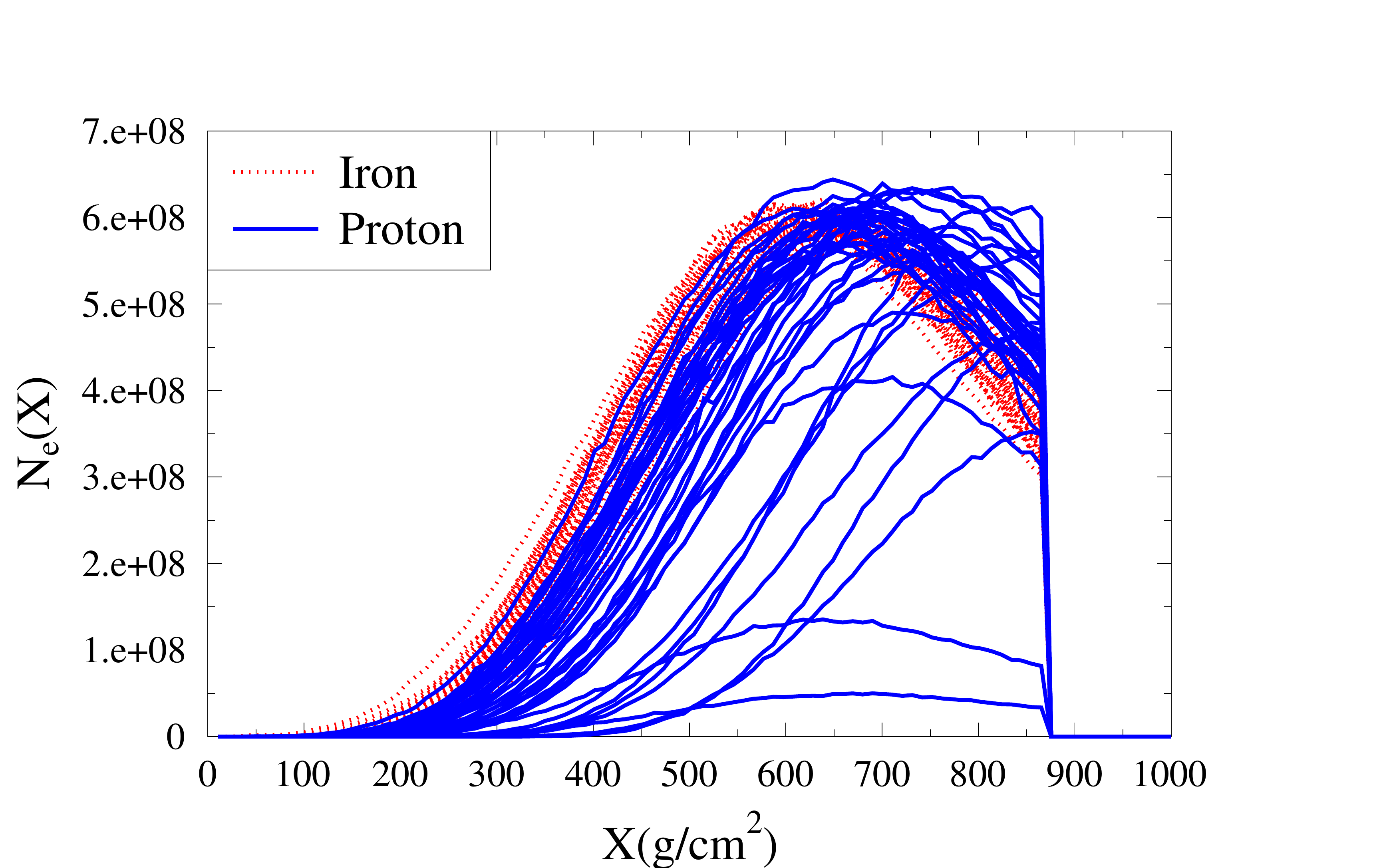}
}
\caption{[color online] Total number of electrons and positrons as function of shower depth
for 40 proton (blue curves) and 40 iron (red curves) initiated showers for $E=10^{18}$\,eV. }
\figlab{sh-prof}
\end{figure}

As an example CONEX-MC-GEO, is used to study the differences of the Lateral Distribution Function (LDF) for two types of showers,  proton- and iron-induced. Simulations are performed for 40 proton- and 40 iron-induced
showers. 
The shower profiles show much larger shower-to-shower fluctuations for proton- than for iron-induced showers, see \figref{sh-prof}. As also is expected, the penetration depth, $X_{max}$, is larger for proton than for iron. The simulations are done for the site of the Pierre Auger Observatory where ground level is at $870$\,g~cm$^{-2}$.
The fraction of excess electrons not very dependent on initial energy or composition of the cosmic ray but is slowly increasing with shower depth.
Near the shower maximum the charge excess is close to $23\%$ of the total number of electrons and positrons while at ground level it has increased to $25\%$.
As a measure of the pancake thickness we have calculated the mean distance of the electrons behind the
shower front, $L=\left\langle h \right\rangle$, the mean pancake thickness parameter. This shows that the mean pancake thickness parameter is almost independent of shower height or energy. It depends on the shower type, varying from $L=3.9$\,m for proton-induced air showers to $L=4.3$\,m for iron.

The extracted parameters are used in MGMR calculations.
The maximum of the field strength as function of distance is shown in \figref{sh-ldf_FeH} for an observer placed on the $x$-axis where the charge-excess and geo-magnetic fields interfere constructively. The slope of the LDF for iron-induced showers is less
steep than that for protons. Close to the shower core the pulse structure is determined by the pancake thickness which is the reason for the minor differences between the two shower types at small distances. Only at large distances the pulse structure is determined by the height of the shower maximum.

\begin{figure}[!ht]
\centerline{
\includegraphics[width=.48\textwidth, keepaspectratio]{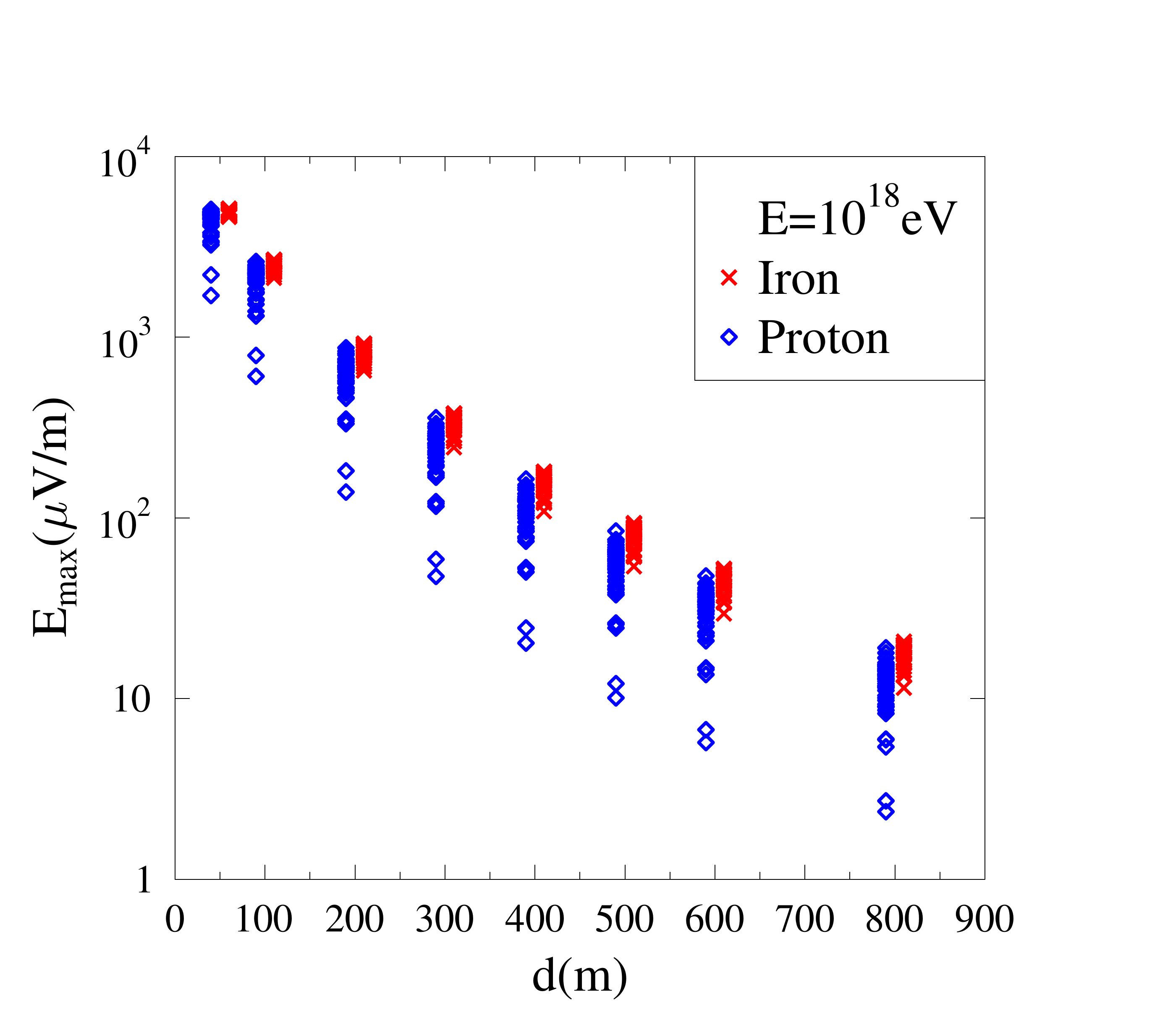}}
\caption{[color online] The LDFs are shown for proton (blue diamonds) and iron (red crosses) induced
showers at an energy of $10^{18}$\,eV, for an observer position at a position where the charge excess and the geo-magnetic contribution interfere constructively.}
\figlab{sh-ldf_FeH}
\end{figure}

In a real experiment the pulse will be filtered in frequency. For this reason we calculated a ratio that
is closer to the data,
\beq
R^{25}_{50/300}=P(50,f>25)/P(300,f>25)
\eeq
where $P(d,f>25)$ is the power in the pulse for frequencies in excess of 25\,MHz at a distance $d$\,m from the shower core.
The value of $R^{25}_{50/300}$ is calculated for each of the simulated proton and iron induced showers separately. Shower-to-shower fluctuations result from the fact that for each of the simulated showers a different profile is extracted. The results are displayed is \figref{sh-pwr} as a histogram. The averages for iron- and proton-induced showers show a large difference, however due to shower-to-shower fluctuations some
of the proton showers result in similar values as iron, directly reflecting their shower profiles.
At an energy of $10^{17}$\,eV the average value of $R^{25}_{50/300}$ differs more. This is a direct
consequence of the fact that the important parameter responsible for the
observed effect, $X_{max}$, differs more.

Due to interference of the charge excess and the geomagnetic contributions the LDF, and thus also
$R^{25}_{50/300}$, depends on the orientation of the observer with respect to the shower axis.
It has been checked that the qualitative dependence on shower type remains as shown
in \figref{sh-pwr}, this because the the charge-excess fraction is similar for the two types of showers.

\begin{figure}[hhc!]
\centerline{
\includegraphics[width=.48\textwidth, keepaspectratio]{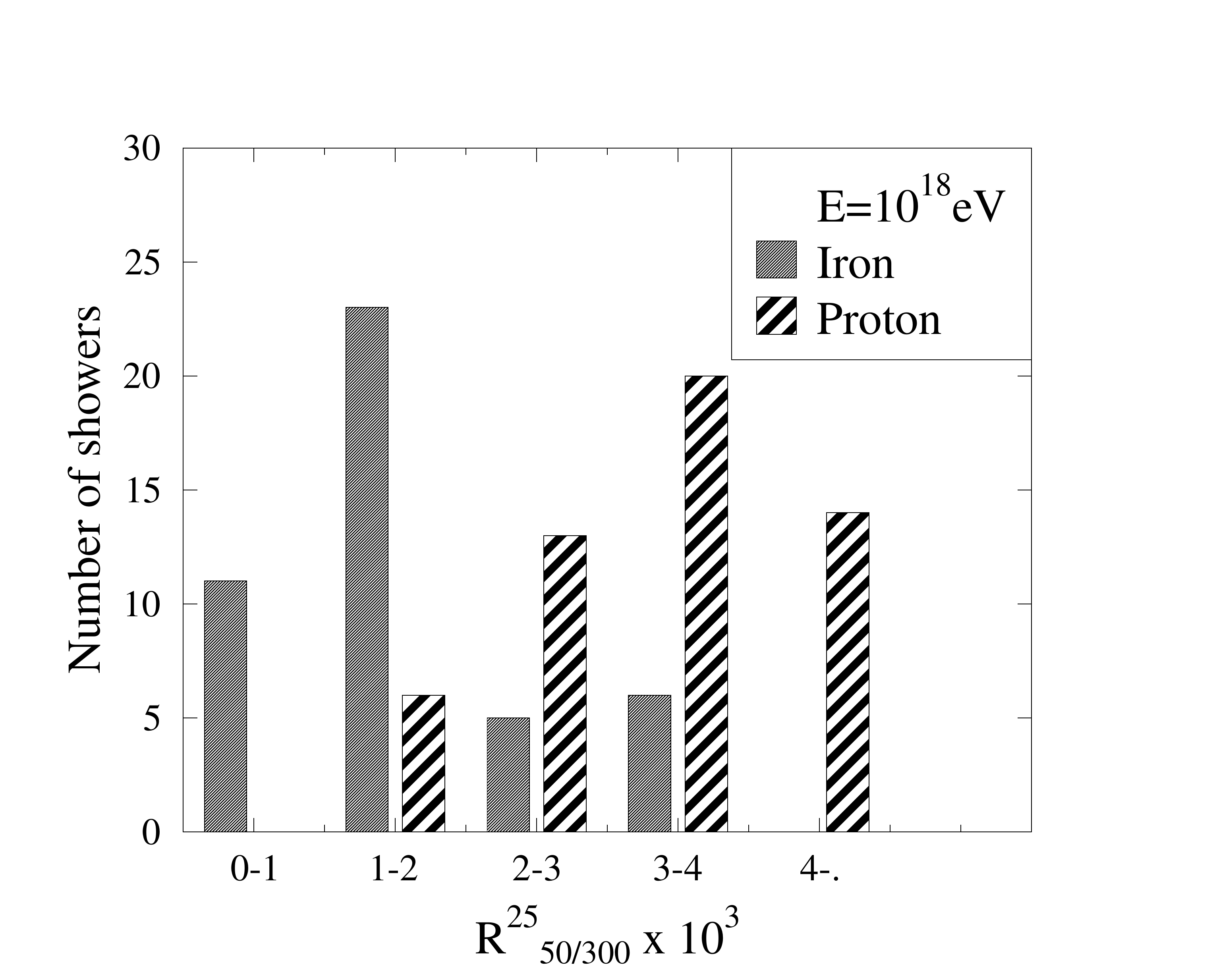}
}
\caption{$R^{25}_{50/300}$, see text, for the 40 simulated proton and iron showers at an energy of $E=10^{18}$\,eV.}
\figlab{sh-pwr}
\end{figure}

\section{Summary and Conclusions}

We have shown that the Macroscopic GeoMagnetic Radiation model for coherent radio emission from extensive air showers offers a very realistic picture for the emitted pulse. It is simple enough to give hands-on insight in the emission mechanism which is of help for the interpretation of the experimental data. Depending on the distance from the shower core there are different length scales that determine the structure of the coherent radio pulse. Close to the shower core the pulse duration is determined by the pancake thickness while at larger distances the shower profile is essential.
This offers the possibility to use the
LDF of the radio pulse to distinguish iron and proton induced showers even though the effects of shower-to-shower fluctuations for protons can be large.

In this presentation little attention has been given to the possibilities to use the polarization of the emitted radiation to disentangle the charge excess and geomagnetic components~\cite{Krijn}. Preliminary experimental results look very promising~\cite{Harm}. Also the effect of the lateral distribution of the particles in the shower and that of a finite index of refraction of air~\cite{Wer08} are being investigated.

\section{Acknowledgment}
This work is part of the research program of the `Stichting voor Fundamenteel
Onderzoek der Materie (FOM)', which is financially supported by the `Nederlandse
Organisatie voor Wetenschappelijk Onderzoek (NWO)'.

\end{document}